\documentclass[12pt,epsf,amstex]{article}
\usepackage [dvips]{graphicx}

\usepackage{amsmath}
\usepackage{amssymb}
\usepackage{amsfonts}
\usepackage{epsfig}
\usepackage{verbatim}
\usepackage{epsfig}
\usepackage{color}

\definecolor{refkey}{rgb}{0,0.5,0.5}
\definecolor{labelkey}{rgb}{0.5,0,0.5}
\definecolor{orange}{rgb}{1.0,0.75,0}

\addtocounter{secnumdepth}{1}
\font\capfont=cmbx12 at 50 pt 
\newbox\capbox \newcount\capl \def\a{A}
\def\docappar{\medbreak\noindent\setbox\capbox\hbox{%
\capfont\a\hskip0.15em}\hangindent=\wd\capbox%
\capl=\ht\capbox\divide\capl by\baselineskip\advance\capl by1%
\hangafter=-\capl%
\hbox{\vbox to8pt{\hbox to0pt{\hss\box\capbox}\vss}}}
\def\cappar{\afterassignment\docappar\noexpand\let\a }

\begin{document}

\newcommand{\ee}{{\rm e}}
\newcommand{\dd}{{\rm d}}
\newcommand{\p}{\partial}
\newcommand{\phX}{\phantom{XX}}

\newcommand{\bc}{\mathbf{c}}
\newcommand{\bC}{\mathbf{C}}
\newcommand{\bq}{\mathbf{q}}
\newcommand{\br}{\mathbf{r}}
\newcommand{\bv}{\mathbf{v}}

\newcommand{\cE}{{E}}
\newcommand{\Ea}{{U}}
\newcommand{\Ed}{{U_2}}
\newcommand{\Et}{{U_3}}

\newcommand{\cH}{{\cal H}}
\newcommand{\Jst}{{J^{\rm st}}}
\newcommand{\cR}{{\cal R}}
\newcommand{\lineJ}{\overline{J}}

\newcommand{\hh}{\frac{1}{2}}
\newcommand{\la}{\langle}
\newcommand{\ra}{\rangle}
\newcommand{\beq}{\begin{equation}}
\newcommand{\eeq}{\end{equation}}
\newcommand{\bea}{\begin{eqnarray}}
\newcommand{\eea}{\end{eqnarray}}
\def\lsim{\:\raisebox{-0.5ex}{$\stackrel{\textstyle<}{\sim}$}\:}
\def\gsim{\:\raisebox{-0.5ex}{$\stackrel{\textstyle>}{\sim}$}\:}

\thispagestyle{empty}
\title{\Large
{\bf A parity breaking Ising chain Hamiltonian\\[4mm] 
as a Brownian motor}\\
\phantom{XXX}
}

\author{{F. Cornu and H.J. Hilhorst}\\[5mm]
{\small Laboratoire de Physique Th\'eorique, b\^atiment 210}\\
{\small CNRS and Universit\'e Paris-Sud,
91405 Orsay Cedex, France}\\}

\maketitle

\begin{small}
\begin{abstract}
\noindent
We consider the translationally invariant but
parity (left-right symmetry) breaking Ising chain Hamiltonian
\begin{equation}
{\cal H} =  
-\Ed\sum_{k} s_{k}s_{k+1} - \Et\sum_{k} s_{k}s_{k+1}s_{k+3}\nonumber
\end{equation}
and let this system evolve by Kawasaki spin exchange dynamics. 
Monte Carlo simulations show 
that perturbations forcing this system off equilibrium
make it act as a Brownian molecular motor which,
in the lattice gas interpretation, transports
particles along the chain.
We determine the particle current under various different
circumstances, in particular as a function of the ratio $\Et/\Ed$
and of the conserved magnetization $M=\sum_ks_k$.
The symmetry of the $\Et$ term in the Hamiltonian is discussed.\\

\end{abstract}
\end{small}
\vspace{12mm}

\noindent LPT Orsay 14/xx
\newpage


{\small

\cappar In biological cells, 
one important class of molecular motors consists of those
that move in a preferential
direction along microtubules 
and transport a cargo of cellular material. 
A directed current is possible only
due to certain specific properties of the cellular environment.
These motors are of the Brownian type and
the transport is
the result of random fluctuations of, say, the local temperature or 
the concentration of cellular constituents, even though
such fluctuations do not {\it a priori\,} favor 
a preferential direction of transport. 
Indeed, if the cell were in a state of thermodynamic equilibrium, 
random fluctuations could not lead to a systematic current.
This intuitively evident fact is formally a consequence of
the equations of physics obeying
detailed balance, or, at the microscopic level,  time reversal
symmetry (TS).
When the system is forced out of
equilibrium, detailed balance no longer holds,
but even in that case we would still be able
to conclude that the current along a microtubule vanished
on the basis of parity symmetry (PS, left-right symmetry) 
 -- if indeed such a symmetry prevailed.

If however
the chemical structure of the microtubule 
is spatially asymmetric along its axis
-- as we know it is --
{\it and\,} the cell is forced out of
equilibrium even so slightly, then both PS and TS are violated.
Then, unless the system still has another invariance,
we no longer have any symmetry argument
available to conclude that there cannot be a current.
These by now well-known facts have been discussed in several review
articles \cite{HanggiBartussek_96,Astumian_97,%
Julicher_etal_97,%
AstumianHanggi_02,%
ReimannHanggi_02,Reimann_02,Linke_02,%
Hanggi_etal_05,%
HanggiMarchesoni_09,%
Kolomeiski_13}.
\vspace{2mm}

It is important to have at hand a diversity of toy models which,
without aiming at direct applicability to specific experimental situations,
illustrate the physical principles that are at work.
Existing models actually show that in the absence of both TS and PS
random fluctuations are rectified and lead to nonzero currents.
In these models TS breaking is easily implemented in a variety of ways, 
such as by submitting the system to a periodic or a random temperature
variation, or to almost any other parity symmetric perturbation.
PS may be also violated in several different manners.

The probably most popular model consists of a  
Fokker--Planck equation for a Brownian 
particle (or: `motor') in a periodic
parity breaking potential, {\it e.g.} an asymmetric sawtooth
potential or Bartussek's double-sine potential \cite{Bartussek_etal_94}. 
Lee {\it et al.} \cite{Lee_etal_03}
exhibit the rectification effect in what is perhaps its simplest form,
by considering 
a particle hopping in discrete time between the nearest neighbor sites
of a one-dimensional lattice
with hopping probabilities that are $3$-periodic 
and asymmetric in space.
Kafri {\it et al.} \cite{Kafri_etal_04} 
derived, with great care for biophysical considerations,
a $2$-periodic lattice model with asymmetric
hops governed by a master equation; this model became popular
among later workers who investigated more formal properties 
involving out-of-equilibrium fluctuations
\cite{Lacoste_etal_08,Verley_etal_11a,Verley_etal_11b}.
Van den Broeck {\it et al.} \cite{vandenBroeck_etal_05} have
studied a model without any periodicity, consisting
of asymmetric triangle-shaped particles in continuous space and
confined to a tube. 
In this case the asymmetry of the interaction leads to 
a nonzero particle current.
\vspace{2mm}

Numerous
physical and chemical properties of macromolecules 
have been reformulated in terms of the dynamics of Ising chains.
Here we add to this list an important phenomenon
that was still missing:
an illustration of the principles of the Brownian molecular motor
entirely in terms of Ising language.
We believe this work is of interest, on the one hand, as a toy model
in the field of Brownian motors, and on the other hand as a
contribution of a new kind to the rich field of Ising model dynamics.
\vspace{2mm}

Let an Ising chain with spin configuration $s=(s_1, s_2,\ldots, s_L)$ 
have the dimensionless Hamiltonian ${\cal H}(s)$ given by
\beq
{\cal H}(s) =  -\Ed\sum_{k=1}^L s_{k}s_{k+1} 
           - \Et\sum_{k=1}^L s_{k}s_{k+1}s_{k+3}\,, 
\label{xHam}
\eeq
where $s_k=\pm1$ and
periodic boundary conditions $s_{L+k}=s_k$ are understood.
The $\Ed$ term in (\ref{xHam}) is the standard Hamiltonian of the
Ising chain with nearest neighbor interactions; the $\Et$ term, 
however, is a novelty:
apart from breaking the up-down spin
symmetry, it also breaks PS
due to the absence of the spin with index $k+2$.
It is, moreover, the simplest possible translationally invariant
term that does so.
We are not aware of PS violating Ising models
ever to have been considered before.
Since the system breaks parity, the possibility -- under appropriate
dynamics -- of a current
along the chain should certainly be expected.
This work is motivated, essentially, by our interest in the
dynamical consequences of the $\Et$ interaction.
The reason for the additional presence of the parity-symmetric $\Ed$ term
will become clear below.

We consider spins up (spins down) as particles (vacancies)
and endow $\cH(s)$ with Kawasaki nearest-neighbor
spin exchange dynamics; each exchange is therefore equivalent to
a particle hopping to an empty nearest-neighbor site.
The total magnetization $M=\sum_ks_k$ or,
equivalently, the total number of particles $N=\tfrac{1}{2}(M+L)$, 
is conserved. 
When the system is in contact
with a heat bath of dimensionless temperature $T$,
we adopt an exchange probability $w_T^{(k,k+1)}$
for the spins $s_k$ and $s_{k+1}$
that satisfies detailed balance with
respect to $\exp(-{\cal H}/T)$. Explicitly,
\beq
w_T^{(k,k+1)}(s) = \frac{1}{2}
\Big[ 1-\tanh\left( {\Delta E^{(k,k+1)}(s)}/{2T} \right) \Big],
\label{xw}
\eeq
where $\Delta E^{(k,k+1)}$ is the energy increment 
associated with the exchange.
Monte Carlo simulations are carried out in the usual way:
each update attempt consists of selecting randomly a link $(k,k+1)$
and exchanging the values of $s_k$ and $s_{k+1}$, if different,
with probability $w_T^{(k,k+1)}$.
The physical time variable $t$ is measured
in units of $L$ update attempts.
\vspace{2mm}

We pause here to note that our approach 
-- the parity breaking Hamiltonian (\ref{xHam}) with the transition
probabilities (\ref{xw}) --
is in some sense opposite to the extensive literature
on kinetically constrained Ising models. 
There one considers a spin system that is either noninteracting or
has a standard parity symmetric Ising Hamiltonian;
and then sets a (possibly PS breaking)
subclass of the usual transition probabilities equal to zero. 
Since the 1980's such models have attracted the interest of physicists and
mathematicians alike as models of glassy relaxation
(see, {\it e.g.} \cite{Palmer_etal_84,%
JackleEisinger_91,SollichEvans_99,%
AldousDiaconis_02,SollichEvans_03,%
Faggionato_etal_12,Chleboun_etal_13}).
Attention to the connection between kinetically constrained
models and certain properties of Brownian molecular 
motors is only very
recent \cite{FaggionatoSilvestri_14a,FaggionatoSilvestri_14b}. 
\vspace{2mm}

We have performed four different sets of simulations on
a system of $L=7\times 10^4$ sites under different
conditions. In order to investigate how the observed phenomena  
depend on the ratio of the interaction constants $\Ed$ and $\Et$
we set
\beq\
\Ed = \sqrt{2}\cos\phi, \qquad \Et = \sqrt{2}\sin\phi,
\label{xE2E3}
\eeq
so that by varying $\phi$ we move along a circle in the $\Ed\Et$ plane.
We set $M=0$ (half-filling) unless stated otherwise.
Each of our simulations started with a fully disordered spin configuration.
\vspace{2mm}

In the first set of simulations we submitted the system to a stepwise varying
temperature of period $\tau=\tau_1+\tau_2$ such that 
\bea
T = T_1, &&\qquad      0 \leq t\mbox{\,mod\,}\tau <\tau_1, 
\nonumber\\
T = T_2, &&\qquad \tau_1 \leq t\mbox{\,mod\,}\tau <\tau.
\eea
In the simulations we took temperatures $T_1=1$ and $T_2=\infty$
and time intervals $\tau_1=2.4$ and $\tau_2=1.6$, whence $\tau=4$.
We set the interaction energies equal to $\Ed=\Et=1$, 
which corresponds to the choice $\phi=\pi/4$ in (\ref{xE2E3}). 
After a transient 
the system properties enter a periodic cycle.
Each period has been divided into $10^3$
subintervals of duration $\tau/1000$. 
After discarding the first $500$ periods we obtained
the average energy per site $E(t)$
and the average particle current per unit time $J(t)$ 
as averages over corresponding subintervals of the next $500$ periods. 
The value of $J(t)$ includes an averaging over all bonds. 
The results are shown in Fig.~\ref{fig_periodic}.

After each temperature step the average
energy per site $\cE(t)$ is seen to start relaxing
towards an equilibrium value at alternatingly $T_1$ and $T_2$.
In the $\tau_1$ phase the net particle current $J(t)$ 
appears to be small, of the order of $0.01$ particle per unit of time.
It is nevertheless clearly different from zero, in contradistinction to
what happens in the infinite temperature $\tau_2$ phase, 
where the Hamiltonian no longer intervenes and parity symmetry dictates
$J(t)=0$ up to fluctuations.

The particle current $\lineJ$ averaged over the full period of
duration $\tau$ is equal to 
$\lineJ=0.0060$ particles per unit of time,
indicated by the dashed horizontal line in Fig.~\ref{fig_periodic}.
This nonzero period averaged current shows that this system acts as a
Brownian molecular motor, which 
is the effect that we wished to demonstrate.
\vspace{2mm}

\begin{figure}
\begin{center}
\scalebox{.30}
{\includegraphics{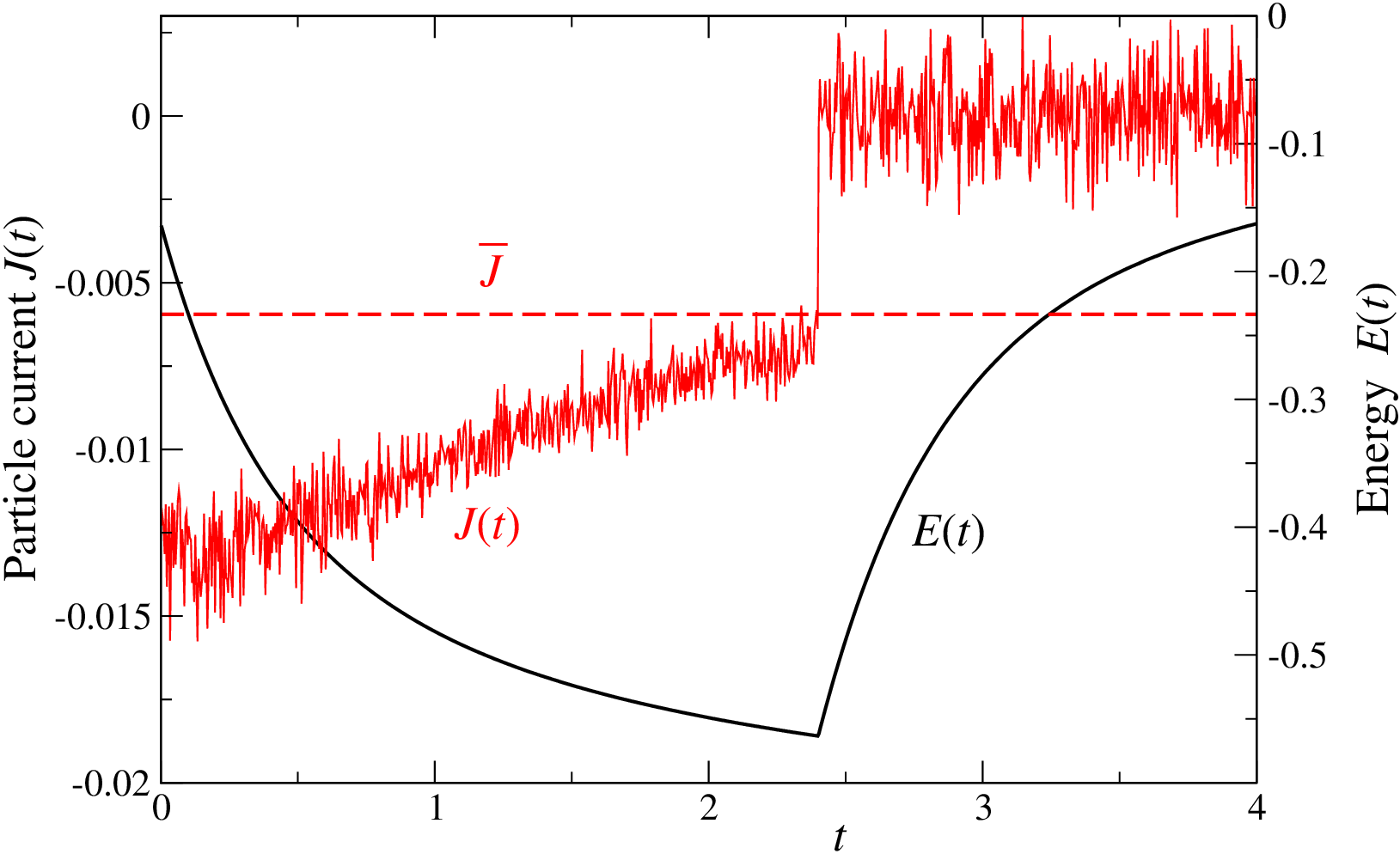}} 
\end{center}
\caption{\small Variation of the average energy per site $\cE(t)$ 
(dimensionless) and particle current $J(t)$ 
(in particles per unit of time)
under a periodic temperature variation with period $\tau=4$.
During the intervals $\tau_1=2.4$ and $\tau_2=1.6$ the temperature is
equal to $T_1=1$ and $T_2=\infty$, respectively.
Parameters are $\Ed=\Et=1$ and $M=0$.
The horizontal dashed line indicates the period-averaged value 
$\bar{J}$ of the current.
}   
\label{fig_periodic}
\end{figure}

In a second set of simulations we 
introduced a `mixing' parameter $0\leq\lambda\leq 1$.
In each update attempt
we submitted the selected spin pair randomly
to the exchange probability $w_{T_1}$ (with probability $1-\lambda$)
or to $w_{T_2}$ (with probability $\lambda$),
which amounts to an effective exchange probability $w$ 
$w=(1-\lambda)w_{T_1}+\lambda w_{T_2}$.
Again we chose $T_1=1$ and $T_2=\infty$ and restricted ourselves to $M=0$.
Under this algorithm, a spin exchange (particle move)
along some bond $(k,k+1)$ which is realized due to $w_{T_2}$ 
may be interpreted as a random energy deposition on that bond.
The system enters a stationary state
which, except for $\lambda=0$ and $\lambda=1$, is
a nonequilibrium one.
In Fig.~\ref{fig_stationary} we show 
the average particle current $\Jst$ as a function of $\lambda$ for 
two different Hamiltonians, {\it viz.}~with $\phi=7\pi/8$ and $\phi=5\pi/8$,
which have $(\Ed,\Et)=(-0.54,1.30)$ and $(\Ed,\Et)=(-1.30,0.54)$, 
respectively. 
In both systems the currents vanish for $\lambda=0$ (equilibrium
at $T_1=1$) and for $\lambda=1$ (equilibrium at $T_2=\infty$), in
agreement with theory, 
but are nonzero in the interior of the
$\lambda$ interval, as expected.
The current is negative for $\phi=7\pi/8$, it changes sign (and is
more than ten times smaller) for $\phi=5\pi/8$, where
it has a shallow minimum of depth $\Jst(5\pi/8)\approx 9.2\times 10^{-5}$
at $\lambda\approx 0.8$.

Each data point in Fig.~\ref{fig_stationary}
was obtained after discarding a transient of
at least $200$ time units before data taking. 
Each data point for $\phi={5\pi}/{8}$
results from an average over $10^4$ time units, except those
in the interval $0.65\leq\lambda\leq 1.00$, where each
results from an average over $2\times 10^5$ time units, this in order
to clearly distinguish it from zero.
For $\phi=7\pi/8$ the error bars are of the order of the symbol sizes;
for $\phi=5\pi/8$ and $0\leq\lambda\leq 0.60$ they are slightly larger
and for $0.65\leq\lambda\leq 1$ they are smaller than the symbol sizes.
\vspace{2mm}

\begin{figure}
\begin{center}
\scalebox{.30}
{\includegraphics{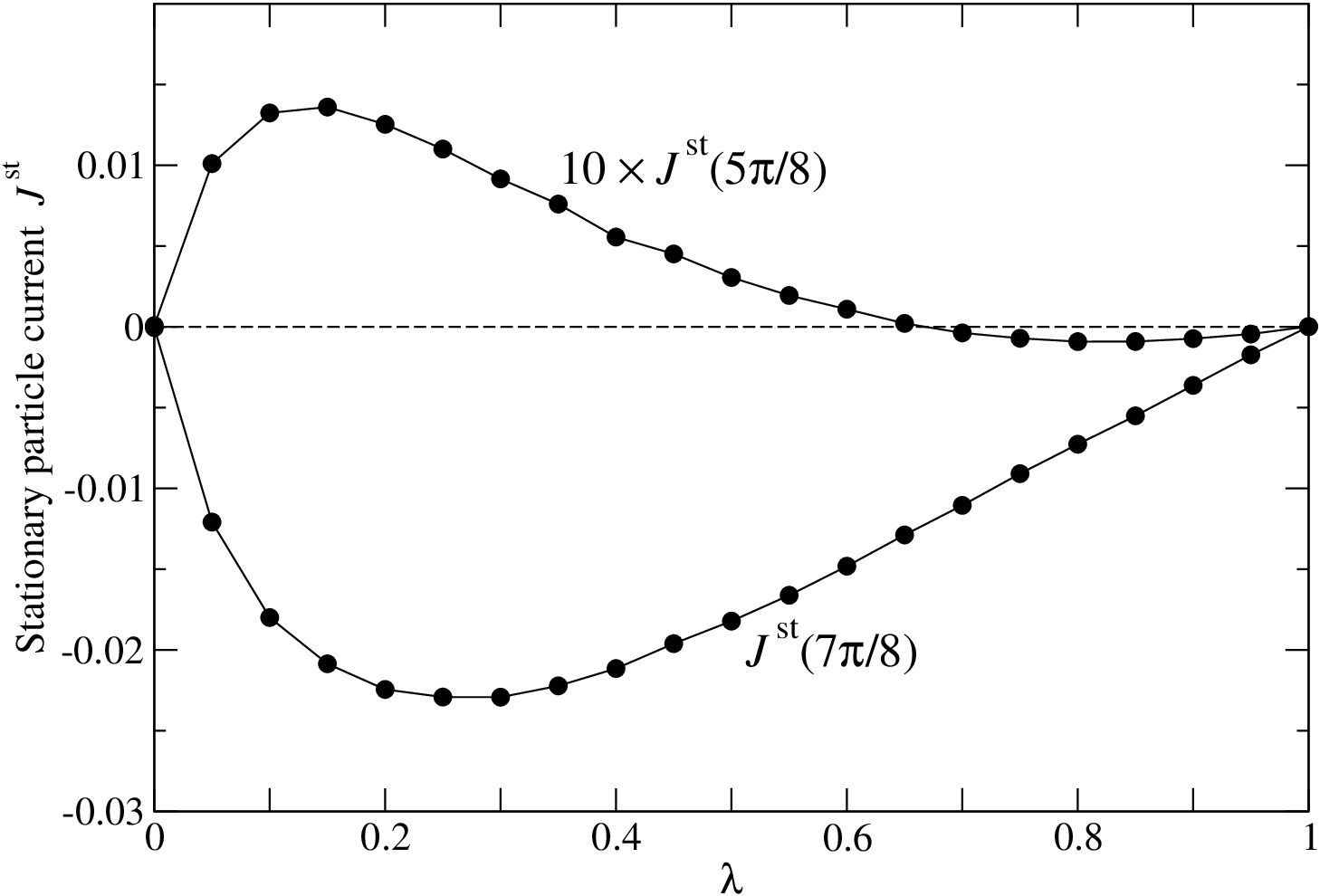}} 
\end{center}
\caption{\small Stationary state particle current $\Jst$ 
for two different values $\phi={5\pi}/{8}$ and $\phi={7\pi}/{8}$
[see Eqs.~(\ref{xHam}) and (\ref{xE2E3})]
at magnetization $M=0$ as a function of the mixing parameter $\lambda$.
This parameter interpolates the transition probabilities between
those of the equilibria at $T_1=1$ (for $\lambda=0$) and
at $T_2=\infty$ (for $\lambda=1$). 
Lines are a guide to the eye.
}   
\label{fig_stationary}
\end{figure}

In a third set of simulations 
we investigated how for fixed $\lambda=0.5$ the average current $\Jst$
depends on the angle $\phi$.
The results are shown in Fig.~\ref{fig_panorama}. 
The symmetry property $\Jst(\phi)=-\Jst(2\pi-\phi)$ and its consequence
$\Jst(\pi)=0$ are easily derived
theoretically and are well borne out by
the simulation. 
When $\phi=0,\pi,2\pi$ the triplet coupling $\Et$ vanishes and the
chain ceases to be parity breaking; in accordance with this the
current vanishes at these points.
The figure shows that even on the
restricted interval $0<\phi<\pi$ the current does not have a unique
sign. There is a flat maximum of height $\Jst\approx 0.0005$
at $\phi\approx 0.59\pi$. 
It is noteworthy that the curve does {\it not\,} pass through
zero at $\phi=\pi/2$ and $\phi=3\pi/2$, as attested by the insert.
We have in fact the extremely small value 
$\Jst(\pi/2)= (2.2\pm 0.3)\times 10^{-5}$.
We will return to this point later.

Each data point in the insert results from an average over 
$2\times 10^6$ time units and its error bar has been indicated; 
each other data point for 
$5\pi/16 \leq \phi \leq 11\pi/16$ results from an average over
$2\times 10^5$ time units and its error bar is smaller than the
symbol size;
and each remaining point in the graph results from an average over
$2\times 10^4$ time units, leading to an error bar of at most the
symbol size.

We remark that
the values of $\phi$ selected for Fig.~\ref{fig_stationary}
are those near the maximum and near the minimum of $\Jst(\phi)$ 
in the interval $0<\phi<\pi$
in Fig.~\ref{fig_panorama}.
\vspace{2mm}

\begin{figure}
\begin{center}
\scalebox{.30}
{\includegraphics{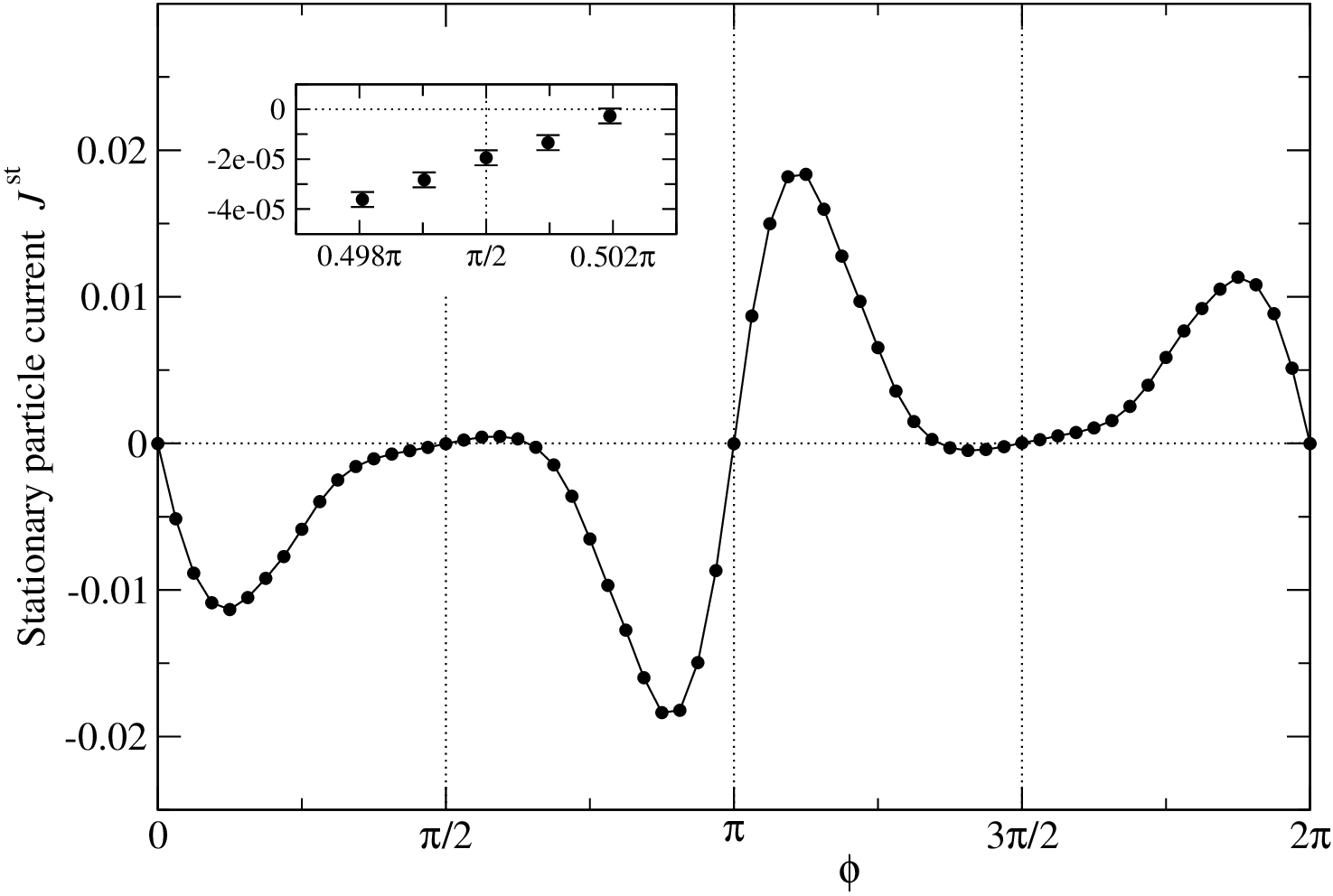}} 
\end{center}
\caption{\small Stationary state particle current $\Jst$ 
for $\lambda=0.5$ and $M=0$
as a function of the angle $\phi$ [Eq.~(\ref{xE2E3})]. 
Parameters are $T_1=1$ and $T_2=\infty$.
Lines are a guide to the eye.
The current is antisymmetric with respect to $\phi=\pi$.
Insert: The curve does {\it not\,} pass through zero at $\phi=\pi/2$. 
}   
\label{fig_panorama}
\end{figure}

The dynamics conserves the number of particles or, in spin language, 
the magnetization. 
All the above simulations were carried out at half-filling, that is,
for a total magnetization $M=0$. We have in the fourth and last
set of simulations considered the pure
triplet Hamiltonian and studied its behavior as a function of 
the magnetization per spin $m=M/L$, or equivalently, the particle
density $\rho = \tfrac{1}{2}(1+m)$.
The results are shown in Fig.~\ref{fig_triplet}.
The current appears to be small for all $m$.
For $m=0$ the same extremely small value of $\Jst(\pi/2)$ 
reappears that was
found in Fig.~\ref{fig_panorama}.
Each of the five data points in the insert is based on an average over
$10^6$ time units; its error bar has been indicated. Each remaining
data point in Fig.~\ref{fig_triplet} 
results from an average over $2\times 10^5$ time
units; its error bar is at most twice the symbol size.
Finally, in simulations for system sizes $L=70$, $700$, and
$7000$, we found within error bars the same value of $\Jst(\pi/2)$, 
so that we may conclude that this nonvanishing current 
is not a finite size effect.
\vspace{2mm}

\begin{figure}
\begin{center}
\scalebox{.30}
{\includegraphics{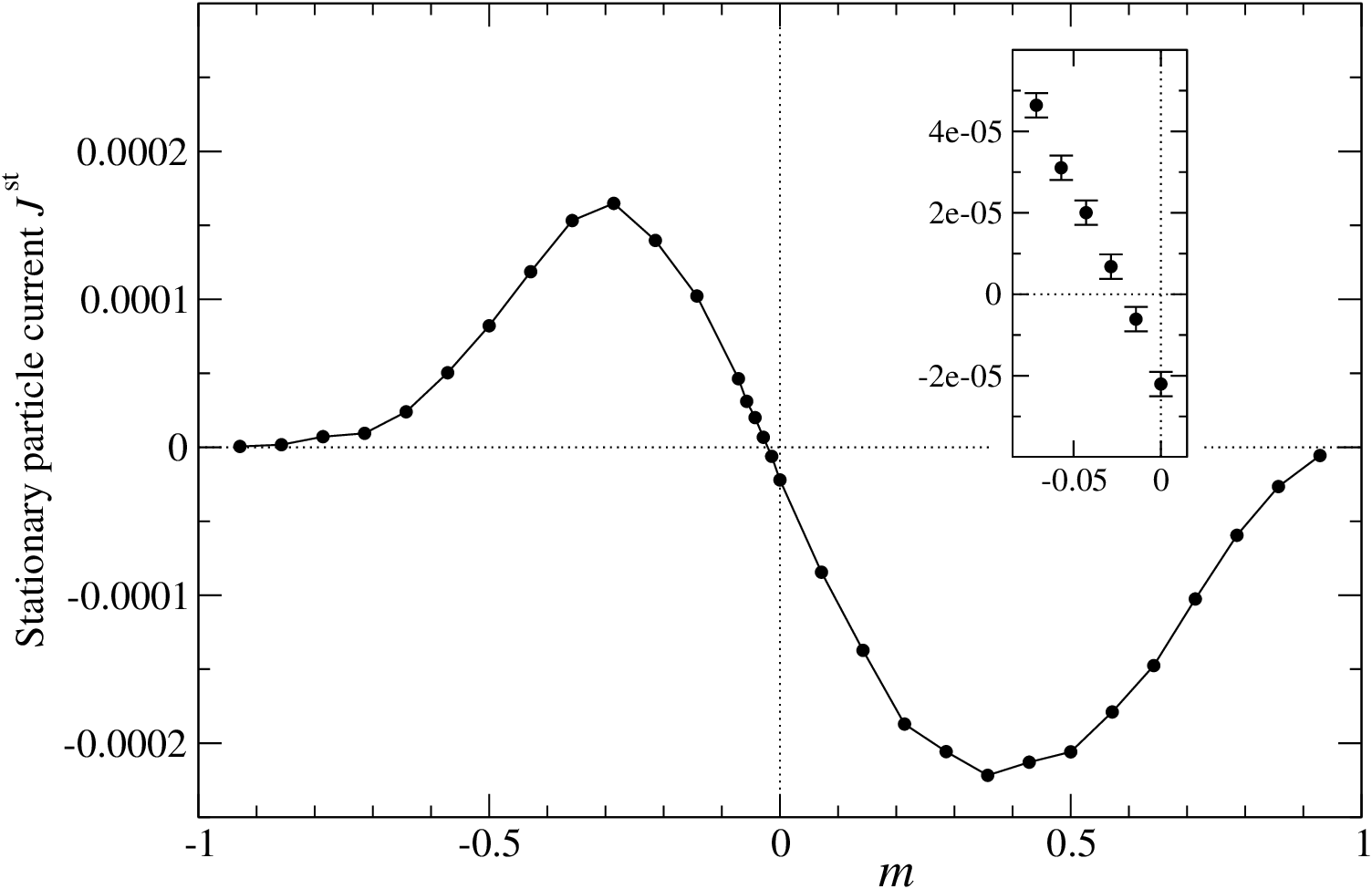}}
\end{center}
\caption{\small Stationary state particle current $\Jst$
for the pure triplet model $\phi = \pi/2$ ({\it i.e.}~$\Ed=0$ and
$\Et=1$) at $\lambda=0.5$
as a function of the magnetization per spin $m=M/L$.
Parameters are $T_1=1$ and $T_2=\infty$.
Lines are a guide to the eye.
The current is very small for all $m$ and does not pass through zero
for $m=0$. 
Insert: Zoom on the five data points closest to the origin.
}   
\label{fig_triplet}
\end{figure}

An initial exploratory simulation for pure triplet
interactions ($\Ed=0$, hence $\phi=\pi/2$) at $M=0$, 
seemed to suggest that there was no motor effect,
and this motivated us to extend our search
by adding to the Hamiltonian the $\Ed$ term.
Subsequent closer inspection has shown, however, that $\Jst(\pi/2)<0$,
as is clearly brought out by the inserts of 
Figs.~\ref{fig_panorama} and \ref{fig_triplet}.
We therefore discuss this case in greater detail.
For the special value $\phi=\pi/2$ the Hamiltonian ${\cal H}$ 
is invariant under each of the transformations 
$\cR_1,\cR_2,\ldots,\cR_7$
defined by $\cR_\ell f(s)=f(s^\ell)$, where
\beq
s^\ell_k = \left\{
\begin{array}{ll}
-s_k & \mbox{if } k\mbox{\,mod\,}7 = 
           (\ell+i)\mbox{\,mod\,}7 \\
     & 
\phantom{XX} \mbox{with } i=0,2,3,\mbox{ or } 4; \\[2mm]
\phantom{-}s_k & \mbox{otherwise}.
\end{array}
\right.
\label{defcRell}
\eeq
We will refer to this additional invariance as `R-symmetry' (RS).
Together with the unit operator, the 
$\cR_\ell$ constitute a group isomorphic to 
$\mathbb{Z}_2\times\mathbb{Z}_2\times\mathbb{Z}_2$.
Hence $\cH$ has eight ground states related by these symmetries:
for $\Et>0$ these are the ferromagnetic state and the seven
states obtained by 
periodically repeating the sequence 
$/++-+---/$ or one of its cyclic permutations.
We observe for later reference that knowing the triplet of spin values 
at three successive positions, say $(s_{j-2},s_{j-1},s_{j})$, 
selects a unique ground state among the eight
(note that the same triplet at another position $j'$
may correspond to a different ground state).
An ensemble $P(s,t)$ in configuration space is R-invariant if
$P(s^\ell,t)=P(s,t)$ for all $\ell=1,2,\ldots,7$. Such an ensemble has
$\langle s_k\rangle=0$ and therefore an average total magnetization
$\langle M\rangle=0$; furthermore, it is easily
shown that under Kawasaki dynamics
it also has an average particle current
$J(t)=0$.

Our simulations are carried out in an $M$-ensemble, that is, one in
which all spin configurations $s$ have exactly $\sum_{k=1}^Ls_k=M$.
Since R-invariance implies that $\langle s_k\rangle=0$,
only a $0$-ensemble can be R-invariant.
However, since the time evolution operator does not commute with the symmetry
operations $\cR_\ell$, a $0$-ensemble which is R-invariant at some
time $t$ will nevertheless
break RS at later times.
We suspect that these facts are responsible, in a way still to be 
elucidated, for the extreme smallness of
$\Jst$ in the case of pure triplet interactions and $M=0$.

The appearance of a 
nonzero current $\Jst$ is 
due to the parity breaking
$\Et$ term in the Hamiltonian (\ref{xHam}). Fig.~\ref{fig_triplet}
shows that (away from the special point $m=0$) this current is typically
of the order $10^{-4}$.
However, comparison of Figs.~\ref{fig_triplet} and \ref{fig_panorama} 
shows that in the presence of the parity {\it symmetric\,} 
$\Ed$ term in the Hamiltonian, $\Jst$ is hugely enhanced
and typically two orders of magnitude larger.
A theoretical explanation of these facts, 
and of the role of RS and its violation, 
remains to be given.
\vspace{2mm}

We make several further comments.
One is best illustrated for the special case $\Et=0$.
Any spin configuration $s$ may then be viewed
as a succession of domains 
that each belong to one of the eight ground states.
The local ground state at site $j$ may be fixed by convention,
{\it e.g.,} as the one belonging to
the spin triplet $(s_{j-2},s_{j-1},s_{j})$.
Domains may therefore have any length $\geq 1$
and are separated by domain walls that cost an energy $2\Et$.
In contrast to what happens in for example a one-dimensional 
$8$-state Potts model with nearest neighbor interactions, certain domain
wall moves require energy barriers to be overcome. 
Due to this thermally activated domain wall motion, the system when
its temperature is lowered sufficiently
will acquire spin glass-like properties.
For $\Ed\neq 0$ the analysis is more complicated but similar
energy barriers exist. 
These properties have no direct bearing on the motor 
effect that we wanted to demonstrate here, but gain importance
when lower temperatures are considered than
those of the present work.
\vspace{2mm}

The model of this paper belongs to the class that incorporates
interactions between the particles (or `motors').
Such collective dynamics of motors enjoys an increasing interest and was 
recently considered by several authors
\cite{PinkoviezkyGov_13,Neri_etal_13,Teimouri_etal_14}.
Interactions are also a feature of
`molecular spiders' \cite{Antal_etal_07}.
A model much more elaborate than ours and simulated 
by Malgaretti {\it et.~al.} \cite{Malgaretti_etal_12},
incorporates both an external potential and 
hydrodynamic interactions between the particles.
Competing teams of motors coupled to the
same cargo and allowing for bidirectional transport
were modeled very recently by
Klein {\it et.~al.} \cite{Klein_etal_14} with great attention to
the values of the biophysical parameters.

The particle interaction due to Hamiltonian (\ref{xHam})
may easily be reexpressed in terms of the site occupation numbers 
$n_k=\tfrac{1}{2}(1+s_k)=0,1$. It involves $1$-, $2$-, and
$3$-particle terms and has a range of
three lattice units. Beyond the hard-core repulsion,
this interaction is, at least for $\Ed,\Et>0$,
essentially attractive:
in order to minimize the energy, particles prefer to 
cluster together.
The precise form of these interactions was of
course motivated only by the simplicity of expression (\ref{xHam}) rather
than by biophysical considerations;
nevertheless, experiments \cite{Roos_etal_08}
seem to suggest that motor proteins interact locally
via short-range potentials that are weakly
attractive and lead to `motor clustering'.

For the toy model of this work
one easily shows that at low (high) particle density
the molecular motor effect requires the presence of at least 
three particles (vacancies) in the same neighborhood.
We therefore 
deduce, 
but have not verified by simulation, that the current 
behaves as $\Jst \sim \rho^3$ in the limit 
$\rho\to 0$\, ($m\to-1$)
and as $\Jst \sim (1-\rho)^3$ for $\rho\to 1$
($m\to 1$).
The curve of Fig.~\ref{fig_triplet}
seems compatible with both relations.

We have seen that here, as in other models,
the direction of the particle current is not easily
predictable {\it a priori,}
and is certainly not intuitive. 
Our simulations show that it is a complicated
function of the model parameters.
\vspace{2mm}


In conclusion, we have for the first time in this work investigated
a PS violating Ising Hamiltonian. 
We have added to the standard nearest neighbor Hamiltonian
the simplest possible of all parity breaking interactions,
{\it viz.} a triplet term. 
We have shown that under Kawasaki
dynamics, and when the system is pulled off equilibrium,
a Brownian motor effect arises.
No appeal was made to any concept extraneous to the Ising model.
We have pointed out theoretical questions that arise
and to which we do not have the answers. 
The most intriguing ones have to do with the symmetry of the $\Et$
term in the Hamiltonian.
Many additional questions may of course be asked.
The ability of particle to climb uphill against an external force
is a necessary direct consequence, that we have however not tested
by simulation.
Nor have we considered, for example, the heat flow
through the system, its efficiency, fluctuation theorems, and so on.
We leave all these questions aside in this short note, 
whose sole purpose has been to present the proof-of-principle
of an `Ising motor'.
\vspace{2mm}

}




\begin{thebibliography}{10}


\bibitem{HanggiBartussek_96}
P.~H\"anggi and R.~Bartussek, 
\newblock{\em Lect. Notes Phys.} {\bf 476} (1996) 294.

\bibitem{Astumian_97}
R.D.~Astumian, 
\newblock{\em Science\,} {\bf 276} (1997) 917.

\bibitem{Julicher_etal_97}
F.~J\"ulicher, A.~Ajdari, and J.~Prost,
\newblock{\em Rev.~Mod.~Phys.} {\bf 69} (1997) 1269.

\bibitem{AstumianHanggi_02}
R.D.~Astumian and P.~H\"anggi, 
\newblock{\em Phys. Today\,} {\bf 55} (11) (2002), 33.

\bibitem{ReimannHanggi_02}
P.~Reimann and P.~H\"anggi, 
\newblock {\em Appl. Phys. A\,} {\bf 75} (2002) 169.

\bibitem{Reimann_02}
P.~Reimann, 
\newblock{\em Phys. Rep.} {\bf 361} (2002) 57.

\bibitem{Linke_02}
H.~Linke, 
\newblock{\em Appl. Phys. A\,} {\bf 75} (2002) 167.

\bibitem{Hanggi_etal_05}
P.~H\"anggi, F. Marchesoni, and F. Nori,
\newblock{\em Ann. Phys.} (Leipzig) {\bf 14} (2005) 51.

\bibitem{HanggiMarchesoni_09}
P.~H\"anggi and F. Marchesoni,
\newblock{\em Rev. Mod. Phys.} {\bf 81} (2009) 387.

\bibitem{Kolomeiski_13}
A.B.~Kolomeisky,
\newblock{\em J.~Phys.: Condens.~Matter\,} {\bf 25} (2013) 463101.

\bibitem{Bartussek_etal_94}
R.~Bartussek, P.~H\"anggi, and J.G.~Kissner,
\newblock{\em Europhys. Lett.} {\bf 28} (1994) 459.

\bibitem{Lee_etal_03}
Y.~Lee, A.~Allison, D.~Abbott, and H.E.~Stanley,
\newblock{\em Phys. Rev. Lett.} {\bf 91} (2003) 220601.

\bibitem{Kafri_etal_04}
Y.~Kafri, D.K.~Lubensly, and D.R.~Nelson,
\newblock{\em Biophysical J.} {\bf 86} (2004) 3373.

\bibitem{Lacoste_etal_08}
D.~Lacoste, A.W.C.~Lau, and K. Mallick,
\newblock{\em Phys.~Rev.~E\,} {\bf 78} (2008) 011915.

\bibitem{Verley_etal_11a}
G.~Verley, R.~Ch\'etrite, and D.~Lacoste,
\newblock{\em J.~Stat.~Mech.} (2011) P10025. 

\bibitem{Verley_etal_11b}
G.~Verley, K.~Mallick, and D.~Lacoste.
\newblock{\em Europhys.~Lett.} {\bf 93} (2011) 10002. 


\bibitem{vandenBroeck_etal_05}
C.~van den Broeck, P.~Meurs, and R.~Kawai,
\newblock{\em New J. Physics\,} {\bf 7} (2005) 10.

\bibitem{Palmer_etal_84}
R.G.~Palmer, D.L.~Stein, E.~Abrahams, and P.W.~Anderson, 
\newblock{\em Phys.~Rev.~Lett.} {\bf 53} (1984) 958.

\bibitem{JackleEisinger_91}
J.~J\"ackle and S.~Eisinger,
\newblock{\em Z. Phys B\,} {\bf 84} (1991) 115.

\bibitem{AldousDiaconis_02}
D.~Aldous and P.~Diaconis, 
\newblock{\em J.~Stat.~Phys.} {\bf 107} (2002) 945.

\bibitem{SollichEvans_03}
P.~Sollich and M.R.~Evans,
\newblock{\em Phys.~Rev.~E\,} {\bf 68} (2003) 031504.

\bibitem{SollichEvans_99}
P.~Sollich and M.R.~Evans,
\newblock{\em Phys.~Rev.~Lett.} {\bf 83} (1999) 3238.

\bibitem{Faggionato_etal_12}
A.~Faggionato, F.~Martinelli, C.~Roberto, and C.~Toninelli,
\newblock{\texttt{arXiv:1205.1607v1 [math.PR]}}.

\bibitem{Chleboun_etal_13}
P.~Chleboun, A.~Faggionato, and F.~Martinelli,
\newblock{\em J.~Stat.~Mech.} (2013) L04001.

\bibitem{FaggionatoSilvestri_14a}
A.~Faggionato and V.~Silvestri,
\newblock{\texttt{arXiv:1401.2256v2 [math.PR]}}.

\bibitem{FaggionatoSilvestri_14b}
A.~Faggionato and V.~Silvestri,
\newblock{\texttt{arXiv:1405.1214v3 [math.PR]}}.

\bibitem{Teimouri_etal_14}
H.~Teimouri, A.B.~Kolomeisky, and K.~Mehrabiani,
\newblock{\texttt{arXiv:1401:6445v1 [cond-mat.stat-mech]}}.

\bibitem{PinkoviezkyGov_13}
I.~Pinkoviezky and N.S.~Gov, 
\newblock{\em New J.~Physics\,} {\bf 15} (2013) 025009.

\bibitem{Neri_etal_13}
I.~Neri, N.~Kern, and A.~Parmeggiani,
\newblock{\em New J.~Physics\,} {\bf 15} (2013) 085005.

\bibitem{Antal_etal_07}
T.~Antal, P.L.~Krapivsky, and K.~Mallick,
\newblock{\em J.~Stat.~Mech.} (2007) 08027.

\bibitem{Malgaretti_etal_12}
P.~Malgaretti, I.~Pagonabarraga, and D.~Frenkel,
\newblock{\em Phys.~Rev.~Lett.} {\bf 109} (2012) 168101.

\bibitem{Klein_etal_14}
S.~Klein, C.~Appert-Rolland, and L.~Santen,
\newblock{\texttt{arXiv:1404.3478v2 [physics.bio-ph]}}.

\bibitem{Roos_etal_08}
W.~Roos, O.~Camp\`as, F.~Montel, G.~Woehlke, J.P.~Spatz, P.~Bassereau,
and G.~Cappello,
\newblock{\em Phys.~Biol.} {\bf 5} (2008) 046004.





\end{thebibliography}

\end{document}